\documentclass[twocolumn,showpacs,preprintnumbers,amsmath,amssymb,superscriptaddress]{revtex4}
\usepackage[dvips]{graphicx}
\usepackage{epsfig}
\usepackage{amsmath}
\newcommand{\vi}[1]{\mbox{\boldmath $#1$}}

\begin{document}
\title{Local versus nonlocal $\alpha\alpha$ interactions in $3\alpha$ description of $^{12}$C}
\author{Y. Suzuki}%
\affiliation{Department of Physics, and Graduate School of 
Science and Technology, Niigata University, Niigata
950-2181, Japan} 
\author{H. Matsumura}
\affiliation{Graduate School of Science and Technology, Niigata
University, Niigata 950-2181, Japan}
\author{M. Orabi}
\affiliation{Graduate School of Science and Technology, Niigata
University, Niigata 950-2181, Japan}
\author{Y. Fujiwara}
\affiliation{Department of Physics, Kyoto University, Kyoto 606-8502, Japan}
\author{P. Descouvemont}
\affiliation{Physique Nucl\'eaire Th\'eorique et Physique Math\'ematique, C.P. 229,\\
Universit\'e Libre de Bruxelles (ULB), B 1050 Brussels, Belgium}
\author{M.~Theeten}
\affiliation{Physique Quantique, C.P. 165/82,  
Universit\'e Libre de Bruxelles (ULB), B 1050 Brussels, Belgium}
\affiliation{Physique Nucl\'eaire Th\'eorique et Physique Math\'ematique, C.P. 229,\\
Universit\'e Libre de Bruxelles (ULB), B 1050 Brussels, Belgium}
\author{D. Baye}  
\affiliation{Physique Quantique, C.P. 165/82,  
Universit\'e Libre de Bruxelles (ULB), B 1050 Brussels, Belgium}
\affiliation{Physique Nucl\'eaire Th\'eorique et Physique Math\'ematique, C.P. 229,\\
Universit\'e Libre de Bruxelles (ULB), B 1050 Brussels, Belgium}

\begin{abstract}
Local $\alpha \alpha$ potentials fail to describe $^{12}$C as a 
$3\alpha$ system. Nonlocal $\alpha \alpha$ potentials that 
renormalize the energy-dependent kernel 
of the resonating group method allow interpreting simultaneously 
the ground state and $0^+_2$ resonance of $^{12}$C as $3\alpha$ states. 
A comparison with fully microscopic calculations provides a measure 
of the importance of three-cluster exchanges in those states.
\end{abstract}
\maketitle

The microscopic $3\alpha$ model describes $^{12}$C 
as a 12-nucleon system where the nucleons are grouped 
into three substructures known as $\alpha$ clusters. 
The merit of this microscopic model is that it starts from 
effective two-nucleon potentials and takes account of the 
antisymmetry requirement exactly. 
It provides a satisfactory qualitative description of several 
states of $^{12}$C including the $0^+$ ground state and the 
astrophysically important $0^+_2$ `Hoyle' resonance 
despite that spin-dependent effects are missing 
in the $\alpha$ cluster assumption. 
The first microscopic $3\alpha$ calculation was performed 
within the resonating group method (RGM) \cite{rgm,saito} 
by Kamimura~\cite{Ka-81}. 
This model provides useful wave functions for various applications 
but remains complicated and heavy to handle. 

This difficulty motivates another development of the cluster model, 
a `macroscopic' $3\alpha$ model, involving three structureless 
$\alpha$ bosons interacting through local $\alpha\alpha$ forces 
which reproduce the $\alpha \alpha$ phase shifts. 
However, in spite of about 40 years of 
efforts \cite{VW-71,smirnov,Tu-01,TBD03}, 
this macroscopic treatment still meets serious problems. 
In the 3$\alpha$ system, any local potential whether shallow or deep 
yields very poor results, in disagreement with experiment and
microscopic models. 
The macroscopic $3\alpha$ model though physically appealing 
fails to reproduce even qualitatively the properties of the 
$^{12}$C ground state. 

Initially shallow $\alpha\alpha$ potentials \cite{ab} were used. 
Such potentials fit the phase shifts and do not support bound states 
since $^8$Be is unbound, but lead to an unrealistically weakly bound $^{12}$C. 
The microscopically founded suggestion that potentials should be 
deep \cite{saito,NKK71,bfw} 
does not improve the situation. 
Deep potentials involve additional bound states, called forbidden 
or redundant states, 
whose role is to simulate the effect of Pauli antisymmetrization 
through orthogonality with physical states. 
They improve physical properties of two-body wave functions 
but introduce many unwanted levels in the spectrum of three-body systems
\cite{AFS92}. 
Forbidden states must thus be eliminated and various techniques 
have been developed~\cite{hori.ocm,smirnov,KP-78,Ba-87}. 
Unfortunately, the $^{12}$C ground-state energy is found to strongly depend 
on how accurately the forbidden states of the deep potential are 
eliminated from the configuration space \cite{Tu-01,TBD03}. 
Strangely, the most accurate calculations provide much worse results than 
unconverged ones; they again provide a weakly bound $^{12}$C \cite{TBD03}. 
The origin of this enigmatic behavior has finally been explained by a careful 
consideration of the role of almost-forbidden states in this 3-cluster system 
\cite{removal,fbs}. 

That the $\alpha\alpha$ potential can not be local for describing $^{12}$C 
is not surprising. 
The interaction between composite particles is intrinsically nonlocal 
because of the exchange symmetry of identical constituents. 
Whether nonlocal $\alpha\alpha$ potentials can provide a description 
of $^{12}$C is thus a fundamental issue. 
Such questions have recently been examined for few-nucleon systems 
using nonlocal nucleon-nucleon interactions of various types 
\cite{dole,viviani,takeuchi,triton}. 

A first attempt to use energy-dependent 
nonlocal $\alpha\alpha$ RGM kernels~\cite{rgm,saito} 
has been developed in Ref.~\cite{fujiwara}. 
This interaction contains all antisymmetrization effects 
in the $\alpha\alpha$ system. It accurately describes the $^8$Be 
ground-state resonance and the $\alpha\alpha$ phase shifts. 
The only model assumptions are a simple $(0s)^4$ description of 
the $\alpha$ clusters 
within the translation-invariant harmonic-oscillator (HO) model 
and the use of an effective nucleon-nucleon interaction \cite{TLT77}. 
Employing the resulting nonlocal $\alpha\alpha$ interaction is rendered complicated 
by the existence of an energy dependence and of Pauli-forbidden states. 
Within this semi-microscopic $3\alpha$ model, the difficulties associated with the 
accuracy of the forbidden-state elimination disappear \cite{fujiwara,theeten}. 
Despite a reasonable success for the ground state, 
the problem now is that the $0^+_2$ excited state is not obtained 
simultaneously. 
Moreover the energy dependence of the $\alpha\alpha$ nonlocal 
interaction raises another difficulty. 
This energy is well defined in a two-body system but not in a three-body 
system. Its choice raises another ambiguity in the model~\cite{theeten}. 

The principle of the elimination of the energy dependence from the RGM equation 
is known for a long time \cite{saito,schmid,timm,book}. 
For two-body systems, its interest is mainly academic. 
That three-body calculations should be performed with the complicated 
nonlocal potential 
resulting from this transformation has been suggested several 
times \cite{schmid,book} but its use has never thoroughly been examined  
in view of the many difficulties involved. 
Indeed, not only must these nonlocal potentials be derived but their use
and evaluation of validity simultaneously 
require mastering accurate calculations of macroscopic three-body 
systems with nonlocal forces and of microscopic three-cluster systems for comparison. 
This know-how has been developed in recent years by our groups 
\cite{fujiwara,fbs,svm,2alpha.rgm,TBD05,matsumura,KD-04}. 
The purpose of the present study is to clarify the role of nonlocality in the 
$\alpha\alpha$ potential by comparing the results of its application 
to a three-cluster description of $^{12}$C with those obtained with local potentials 
on one hand and with the fully microscopic $3\alpha$ model on the other hand. 
 
As a microscopic theory for the intercluster potential, 
we start from the RGM \cite{rgm,saito}. 
The RGM wave function of the two-cluster (A+B) system 
with the relative motion function $\chi$ is expressed as 
\begin{equation}
\Psi={\cal A}[\phi_A \phi_B \chi({\vi x})]=\int \chi({\vi
 r})\, \Phi({\vi r})\, d{\vi r},
\end{equation}
with the basis function 
$\Phi({\vi r})={\cal A}[\phi_A \phi_B \delta({\vi x}-{\vi r})]$, where 
$\phi_A$ and $\phi_B$ are the cluster internal (i.e. translation invariant) 
wave functions and ${\cal A}$ is an antisymmetrizer over all nucleons. 
Here ${\vi x}$ is the relative coordinate between the clusters, 
while ${\vi r}$ is a parameter coordinate corresponding to ${\vi x}$. 
The RGM equation for $\chi$ reads
\begin{equation}
(T+V+\varepsilon K)\chi=\varepsilon \chi, \ \ \ \ \ 
V=V_D+V^{EX}.
\label{rgmeq}
\end{equation}
Here $T$ is the intercluster kinetic energy, 
$V_D$ is local and called the direct potential, and 
$V^{EX}=K_T+K_V$, where 
$K_T$ and $K_V$ are the exchange nonlocal kernels for 
the kinetic and potential (including 
the Coulomb term) energies, respectively. 
The norm kernel is defined by 
\begin{equation}
{\cal N}({\vi r},{\vi r}') 
\equiv \langle \Phi({\vi r})\mid \Phi({\vi r}') \rangle 
=\delta({\vi r}-{\vi r}')-K({\vi r},{\vi r}'),
\end{equation}
with the short-range overlap kernel $K$.  
In the $\alpha+\alpha$ case, these kernels are given in Ref.~\cite{theeten}. 
The energy $\varepsilon$ is defined with respect to the $A+B$ threshold. 

Equation~(\ref{rgmeq}) suggests that the $A+B$ potential is 
\begin{equation}
V^{\varepsilon K}=V+\varepsilon K. 
\label{pot.ek}
\end{equation}
When employing this kind of energy-dependent 
RGM kernels in three-cluster systems, 
the $\varepsilon$ value has to be provided for each pair of clusters 
but is not a well defined quantity. 
A self-consistent procedure providing an average value of the energy distribution 
has been proposed in Ref.~\cite{fujiwara}. 
This procedure leads to a reasonable energy for the $^{12}$C ground state 
but fails for the $^9$Be system~\cite{theeten}. 

To resolve this problem one has to obtain a potential which 
has no explicit energy-dependence and still 
maintains basic properties such as the phase shifts. 
Such a potential can be constructed for 
a macroscopic (or renormalized) relative motion 
function, 
\begin{equation}
g=\sqrt{{\cal N}}\chi, 
\end{equation}
by requiring the condition~\cite{timm} 
\begin{equation}
(T+V^{RGM})g=\varepsilon g.
\end{equation}
The nonlocal potential $V^{RGM}$ is expressed as
\begin{equation}
V^{RGM}=V+W,
\end{equation}
where the new nonlocal operator $W$ is the difference between the 
renormalized RGM potential $V^{RGM}$ and the bare RGM potential $V$,
\begin{equation}
W={\cal N}^{-1/2}(T+V){\cal N}^{-1/2}-(T+V).
\end{equation}

The relative motion functions $g$ have the nice property 
$\langle g | g' \rangle \!=\! \langle {\cal A}[\phi_A \phi_B \chi]|
{\cal A}[\phi_A \phi_B \chi'] \rangle$, that is, 
the orthonormality of microscopic wave functions $\Psi$ and $\Psi'$ 
is precisely transmitted to $g$ and $g'$. 
In addition, the asymptotics of $g$ is the same as that of $\chi$ 
because ${\cal N}$ approaches unity at large distances. 
The phase shift determined from $g$ using the potential $V^{RGM}$ is 
exactly equal to that determined from the RGM equation (\ref{rgmeq}) for $\chi$. 

A particular function $\chi_f$ is called a Pauli-forbidden state (PFS) 
if it satisfies the equation 
\begin{equation}
{\cal N}\chi_f=0 {\rm\ \ or\ \ }  K\chi_f=\chi_f,
\end{equation}
which, by using Eq.~(\ref{rgmeq}), leads to $(T+V)\chi_f=0$, 
and also $W\chi_f=0$. 
We thus have the property 
\begin{equation}
(T+V^{RGM})\chi_f=0,
\end{equation}
justifying the denomination of forbidden states. 

More generally, consider the eigenvalue problem 
\begin{equation}
K\psi_{\kappa}=\kappa \psi_{\kappa}.
\label{eigfn.K}
\end{equation}
The PFS are nothing but the eigenfunctions with eigenvalue $\kappa=1$. 
For cluster wave functions described with HO functions with a 
common size parameter, the eigenvalue problem (\ref{eigfn.K}) can be 
solved~\cite{suppl.horiuchi}. 
The eigenfunctions are HO wave functions $\psi_{n\ell m}$, and the 
corresponding eigenvalues are given by 
$\kappa_{n\ell}=4(1/2)^{2n+\ell}-3\delta_{2n+\ell,0}$ 
for $\alpha+\alpha$. 
There are thus three PFS, 
$\psi_{000}, \, \psi_{100}$ and $\psi_{02m}$ for $\ell$ even, and 
all states are PFS for $\ell$ odd. 

The nonlocal kernel $V^{RGM}$ can 
be expanded in terms of the eigenfunctions of $K$, e.g., 
$W$ is expressed as 
\[
W({\vi r} , {\vi r}')={\sum}'_{nn'\ell m}W_{nn'\ell}
\psi_{n\ell m}({\vi r})\psi_{n'\ell m}^*({\vi r}'),
\]
\vspace*{-5mm}
\begin{equation}
W_{nn'\ell}=\Big[\frac{1}{\sqrt{(1-\kappa_{n\ell})(1-\kappa_{n'\ell})}}
-1\Big]\langle \psi_{n\ell m} |T+V| \psi_{n'\ell m} \rangle,
\label{nlpot.exp}
\end{equation}
where the prime indicates that the PFS are excluded from the sum. 
Here $V$ is assumed to be rotation-invariant, which makes its 
matrix element independent of $m$. 
In practice, we include $n$ and $n'$ up to 100. 

The potential $V^{RGM}$ is expanded into partial waves as 
\begin{eqnarray}
V^{RGM}_{\ell}(r,r')=2\pi rr'\int_{-1}^{1}V^{RGM}({\vi r},{\vi r}')
P_{\ell}(t)\, dt  \nonumber \\
=V_D(r)\, \delta(r-r')+V^{EX}_{\ell}(r,r')+W_\ell(r,r')
\end{eqnarray}
with $t= {\vi r}\cdot {\vi r}'/rr'$ and 
$P_{\ell}$ is a Legendre polynomial. 
The potential $V^{RGM}_{\ell}$ acts on $g_{\ell}$, the 
macroscopic wave function in partial wave $\ell$. 
The terms $V_D$ and $V^{EX}_{\ell}$ are the direct and nonlocal RGM 
potentials. 
Using $\psi_{n\ell m}({\vi r})=R_{n\ell}(r)Y_{\ell m}(\hat{\vi r})$ 
enables one to calculate $W_{\ell}$ as 
\begin{equation}
W_{\ell}(r,r')=rr'{\sum}'_{nn'}W_{nn'\ell}R_{n\ell}(r)R_{n'\ell}(r').
\end{equation} 

\begin{figure}[b]
\vspace{-3mm}
\caption{Contour plots of $V^{EX}_{\ell}(r,r')$ (left) and 
$W_{\ell}(r,r')$ (right) in MeV\,fm$^{-1}$ 
for the $\alpha+\alpha$ system. 
The $\ell$ values are 0, 2 and 4 from top to bottom.}
\label{nl.alpha-alpha}
\vspace{-1mm}
\begin{center}
\hspace*{0.3cm}
\resizebox{9cm}{!}{\includegraphics*[2cm,0.1cm][9.3cm,8.5cm]{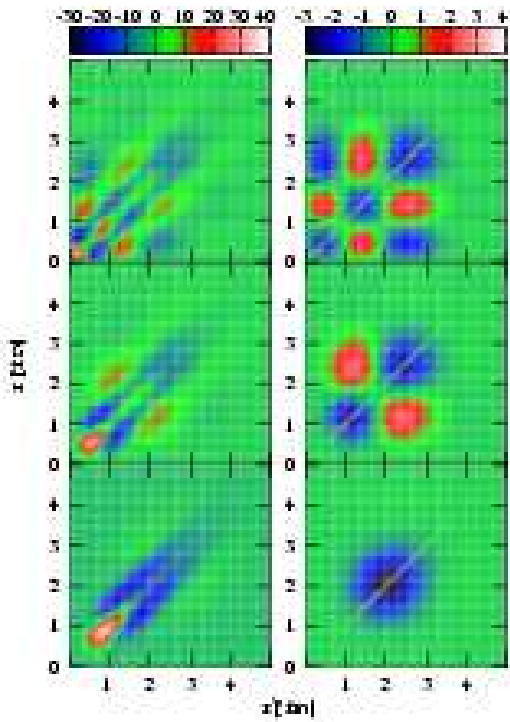}}
\end{center}
\vspace{-8mm}
\end{figure}

Figure~\ref{nl.alpha-alpha} displays $V^{EX}_{\ell}(r,r')$ 
and $W_{\ell}(r,r')$ for the $\alpha+\alpha$ system. 
The parameters of the RGM kernels ($\nu = 0.257$ fm$^{-2}$, $u = 0.94687$) 
taken from Ref.~\cite{2alpha.rgm} reproduce the $\alpha+\alpha$ phase shifts. 
The nonlocality pattern is very different between $V^{EX}_{\ell}$ and $W_{\ell}$. 
The potentials for $\ell=0$ and 2 show a rapidly oscillating behavior, 
which is related to the existence of the PFS. 
The absolute value of $W_{\ell}$ is smaller than that of $V^{EX}_{\ell}$ 
by about one order of magnitude. 
The major contribution to the nonlocal potential comes from the kinetic energy; 
its effect is reduced by the potential energy term. 
\begin{table}[b]
\vspace{-4mm}
\caption{Comparison of the energies (in MeV) from 3$\alpha$ threshold 
and the rms radii (in fm) of point $\alpha$-particle distribution 
for the $3\alpha$ system with different 
models and potentials. The Coulomb potential is included. 
Experimental energies are $-7.27$ and 0.38 MeV 
for the $0^+_1$ and $0^+_2$ states.}
\label{12C.table}
\begin{center}
\begin{tabular}{llccc}
\hline\hline
Model &Potential &$L^\pi$ & $E$   & $\sqrt{\langle r^2\rangle}$ \\ 
\hline
Local     & ABd~\protect\cite{ab} & $0_1^+$  & $-$1.52 & 2.34\\
\cline{2-5}
          & BFW(i)~\protect\cite{bfw} & $0_1^+$  & $-$20.62 & 1.29 \\
&               & $0_2^+$  &  $-$1.25 & 2.34 \\
\cline{2-5}
          & BFW(ii)~\protect\cite{bfw} & $0_1^+$  & $-$0.66 & 2.31 \\
\hline
Semi-micro      & $V$ 
                & $0_1^+$  & $-$19.50  & 1.42\\
                & 
                & $0_2^+$  &  $-$0.77  & 2.49\\
\cline{2-5}
                & $V^{\varepsilon K}$ 
                & $0_1^+$  &  $-$9.60  & 1.45\\
\cline{2-5}
                & $V^{RGM}$ 
                & $0_1^+$  & $-$9.44   & 1.62\\
&               & $0_2^+$  &  0.597    & ---\\
\hline
Micro     &
                & $0_1^+$  & $-$11.37 &  1.64\\
&
                & $0_2^+$  &     0.597 &  ---\\
\hline\hline
\end{tabular}
\end{center}
\vspace{-3mm}
\end{table}

Table~\ref{12C.table} compares energies and  
root mean square (rms) radii of the 3$\alpha$ 
system obtained using local potentials (Local), 
the nonlocal $V^{\varepsilon K}$ and $V^{RGM}$ potentials 
(Semi-micro) as well as with fully microscopic calculations 
(Micro) \cite{matsumura,KD-04}. 
The values $\hbar^2/m_N = 41.47$ MeV fm$^2$ and $m_\alpha = 4 m_N$ 
are used everywhere for consistency between Micro and non-Micro 
calculations. 
Results in Refs.~\cite{TBD03,matsumura} are slightly different 
because of the use of other parameter values. 
The two-nucleon potential used in Micro is the same 
as the one used to derive the RGM kernels, and thus the comparison 
between Micro and Semi-micro indicates how closely the various 
potentials simulate the microscopic 3$\alpha$ calculation. 
We stress that all models, except for $V$ of Semi-micro, 
reproduce the $\alpha\alpha$ phase 
shifts in an essentially identical way.  
The rms radius of the point $\alpha$-particle 
distribution is obtained for Micro  
by subtracting $r_\alpha^2$ ($r_\alpha = 1.479$ fm) from 
the $\langle r^2\rangle$ value calculated at the nucleon level. 
As the 0$^+_2$ state calculated with Micro and $V^{RGM}$ is above the 
3$\alpha$ threshold, the value of its rms radius depends 
on the basis choice. Our
calculations suggest, however, that it is very large around 3.5 fm. 

As explained in Refs.~\cite{removal,fbs,TBD03}, 
the local potentials, shallow (ABd) \cite{ab} and deep (BFW) \cite{bfw}), 
never reproduce the microscopic results. 
They lead to a weakly bound (ground or excited) state 
and, sometimes, to a deeply bound ground state. 
The results with BFW depend on the definition of the PFS~\cite{removal}, 
i.e. whether they are (i) HO states or (ii) bound states of the BFW potential. 
In the latter case, the result is similar to that of 
the ABd case. 
The bare RGM potential $V$ cannot reproduce the microscopic 
results either. 
Its energies resemble those obtained with local potentials. 
Only the Semi-micro models with $V^{\varepsilon K}$ and $V^{RGM}$, 
which take account of the energy-dependent nonlocality, 
give results close to Micro. 
In the $V^{\varepsilon K}$ model, however, 
both the $0^+_1$ and $0^+_2$ states cannot be obtained without either loosing
their mutual orthogonality or loosing the self-consistency of $\varepsilon$. 
Only the $V^{RGM}$ potential simultaneously gives satisfactory results 
for the two $0^+$ states. 
We also stress that the rms radius of Micro is reproduced very 
well by the $V^{RGM}$ model. 
Another nice property of $V^{RGM}$ is that the expectation values of the 
3$\alpha$ kinetic and potential energies are almost equal to the 
corresponding values of Micro, while the other models 
including $V^{\varepsilon K}$ give quite different values. 

Though the Semi-micro model with $V^{RGM}$ is qualitatively successful, 
its ground state energy is about 1.9 MeV higher than Micro. 
The former takes account of binary exchanges between $\alpha$ clusters 
but not of simultaneous exchanges among the three $\alpha$ clusters. 
Such three-cluster exchanges produce attraction as observed here. 
This effect is rather significant in the compact ground state 
but negligible in the spatially extended $0^+_2$ state. 

Similar calculations are in progress for $^9$Be treated as 
$\alpha\alpha n$. A preliminary 
value $-2.16$ MeV obtained with $V^{RGM}$ for its $3/2^-$ ground state 
is close to the microscopic result $-2.61$ MeV \cite{theeten} 
and to the experimental value $-1.57$ MeV. 
The use of $V^{RGM}$ thus resolves the overbinding of the Semi-micro energy 
($-3.86$ MeV) with the energy-dependent $V^{\varepsilon K}$ \cite{theeten}. 

To summarize, while the microscopic $3\alpha$ cluster model provides a simultaneous 
description of both $0^+$ states of $^{12}$C, macroscopic $3\alpha$ models 
with local forces that reproduce the $\alpha \alpha$
phase shifts always fail. 
Therefore the very existence of clusters in the $^{12}$C ground state 
could be doubted. 
We have shown that the introduction of microscopically founded nonlocal 
forces, which act in the Pauli-allowed space, solves this problem. 
The nonlocality arises from two different physical origins, 
the exchange of identical constituent 
particles already known in the nonlocal RGM kernels 
and the elimination of the energy-dependence of these kernels. 
Although the latter term is smaller than the former, its contribution is essential 
to simulate the microscopic features. 
The results of the semi-microscopic and microscopic calculations are 
then for the first 
time qualitatively close and even very similar for the $0^+_2$ state. 
The difference of about 2 MeV for the ground state provides a measure 
of three-cluster exchange effects in this state. 

The cluster-model description of $^{12}$C is approximate mainly by its neglect 
of spin effects. 
Nevertheless it provides tractable wave functions which can be very useful 
in reaction and decay models. 
The semi-microscopic model is interesting as it provides similar wave functions 
in much shorter computing times with the possibility of easily improving them 
by tuning the effective force parameters to accurately reproduce the
binding energies. 

This text presents research results of Bilateral Joint Research
Projects of JSPS (2006-2008), Grant-in-Aid for Scientific
Research from JSPS (No. 18540261), a Grant for Promotion
of Niigata University Research Projects (2005-2007), and 
the IAP program P5/07 
initiated by the Belgian-state 
Federal Services for Scientific, Technical and Cultural Affairs. 
M.T. is supported by a scholarship of IISN, Belgium. 
P.D. acknowledges the support of FNRS, Belgium. 
Y.S. thanks FNRS for supporting his stay at ULB 
in the summer of 2006.


\begin{thebibliography}{99}
\bibitem{rgm} K. Wildermuth and Y.C. Tang, {\it A Unified Theory 
of the Nucleus} (Vieweg, Braunschweig, 1977).
\bibitem{saito} S. Saito, Prog. Theor. Phys. Suppl. {\bf 62}, 11 
(1977). 
\bibitem{Ka-81} M.~Kamimura, Nucl. Phys. {\bf A351}, 456 (1981).
\bibitem{VW-71} J.L. Visschers and R. Van Wageningen, Phys. Lett. {\bf 34B}, 
455 (1971).
\bibitem{smirnov} Yu. F. Smirnov, I. T. Obukhovsky, Yu. M. Tchuvil'sky, and 
V. G. Neudatchin, Nucl. Phys. A {\bf 235}, 289 (1974).
\bibitem{Tu-01} E.M.~Tursunov, J. Phys. G {\bf 27}, 1381 (2001). 
\bibitem{TBD03} E.M.~Tursunov, D.~Baye, and P.~Descouvemont, 
Nucl. Phys. {\bf A723}, 365 (2003). 
\bibitem{ab}S. Ali and A. R. Bodmer, Nucl. Phys. {\bf 80}, 99 (1966).
\bibitem{NKK71} V.G. Neudatchin, V.I. Kukulin, V.L. Korotkikh, and 
V.P. Korennoy, Phys. Lett. {\bf 34B}, 581 (1971).
\bibitem{bfw}B. Buck, H. Friedrich, and C. Wheatley, Nucl. Phys. 
A {\bf 275}, 246 (1977).
\bibitem{AFS92} R.M. Adam, H. Fiedeldey, and S.A. Sofianos, 
 Phys. Lett. B {\bf 284}, 191 (1992).
\bibitem{hori.ocm} H. Horiuchi, Prog. Theor. Phys. {\bf 51}, 1266 (1974). 
\bibitem{KP-78} V.I. Kukulin and V.N. Pomerantsev, Ann. Phys. (N.Y.) 
 111 (1978) 330.
\bibitem{Ba-87} D. Baye, Phys. Rev. Lett. {\bf 58}, 2738 (1987).
\bibitem{removal}H. Matsumura, M. Orabi, Y. Suzuki, and Y. Fujiwara, 
Nucl. Phys. A {\bf 776}, 1 (2006).
\bibitem{fbs} Y. Fujiwara, Y. Suzuki, and M. Kohno, Phys. Rev. C {\bf 69}, 
037002 (2004);  
Few-Body Systems {\bf 34}, 237 (2004).
\bibitem{dole} P. Doleschall, I. Borb\'ely, Z. Papp, and
W. Plessas, Phys. Rev. C {\bf 67}, 064005 (2003).
\bibitem{viviani} M. Viviani {\it et al.}, 
Few-Body Systems {\bf 39}, 159 (2006). 
\bibitem{takeuchi} S. Takeuchi, T. Cheon, and E.F. Redish, Phys. Lett. 
B {\bf 280}, 175 (1992).
\bibitem{triton}Y. Fujiwara {\it et al.}, 
Phys. Rev. C {\bf 66}, 021001(R) (2002).
\bibitem{fujiwara}Y. Fujiwara {\it et al.}, 
Prog. Theor. Phys. {\bf 107}, 745 (2002);  
{\it ibid}. {\bf 107}, 993 (2002).
\bibitem{TLT77} D. R.~Thompson, M.~LeMere, and Y.C.~Tang,
	Nucl. Phys. {\bf A286}, 53 (1977).
\bibitem{theeten}M. Theeten, D. Baye, and P. Descouvemont, Phys. Rev. C
	{\bf 74}, 044304 (2006).
\bibitem{timm} W. Timm, H.R. Fiebig, and H. Friedrich, Phys. Rev C {\bf 25}, 
79 (1982).
\bibitem{book}Y. Suzuki, R.G. Lovas, K. Yabana, and K. Varga, {\it
	Structure and Reactions of Light Exotic Nuclei}, (Taylor \&
	Francis, London, 2003).
\bibitem{schmid}E. W. Schmid, Nucl. Phys. A {\bf 416}, 347c (1984). 
\bibitem{svm} K. Varga and Y. Suzuki, Phys. Rev. C {\bf 52}, 2885 (1995).
\bibitem{matsumura}H. Matsumura and Y. Suzuki, Nucl. Phys. A {\bf 739},
	238 (2004).
\bibitem{2alpha.rgm} Y. Fujiwara {\it et al.}, Phys. Rev. C {\bf 70}, 024002
	(2004). 
\bibitem{TBD05} M.~Theeten, D.~Baye, and P.~Descouvemont, Nucl. Phys. {\bf A753}, 233 (2005).
\bibitem{KD-04} S.~Korennov and P.~Descouvemont, Nucl. Phys. {\bf A740}, 249 (2004).
\bibitem{suppl.horiuchi}H. Horiuchi, Prog. Theor. Phys. Suppl. {\bf 62}, 90 (1977). 
\end{thebibliography}
\end{document}